\documentclass[pdftex,twocolumn,epjc3]{svjour3}          

\RequirePackage[T1]{fontenc}

\smartqed  

\RequirePackage{graphicx}
\RequirePackage{mathptmx}      
\RequirePackage{flushend}
\RequirePackage[numbers,sort&compress]{natbib}
\RequirePackage[colorlinks,citecolor=blue,urlcolor=blue,linkcolor=blue]{hyperref}
\usepackage{amsmath}
\journalname{Eur. Phys. J. C}

\begin{document}

\title{The model-independent degeneracy-breaking point in cosmological models with interacting Dark Energy and Dark Matter}


\author{Z. Zhou\thanksref{addr1}
        \and
        T. J. Zhang\thanksref{e1,addr2,addr3}
        \and
        T. P. Li\thanksref{addr1}
}

\thankstext{e1}{E-mail: tjzhang@bnu.edu.cn}

\institute{Tsinghua Center for Astrophysics, Department of Physics, Tsinghua University\\Haidian District, Beijing 100084, P. R. China\label{addr1}
          \and
          Department of Astronomy, Beijing Normal University\\Haidian District, Beijing 100875, P. R. China\label{addr2} 
\and Institute for Astronomical Science, Dezhou University\\Dezhou 253023, China\label{addr3}}

\date{Received: date / Accepted: date}

\maketitle

\begin{abstract}
We study cosmological models with interaction between dark energy (DE) and dark matter (DM). For the interaction term $Q$ in cosmic evolution equations, there is a model-independent degeneracy-breaking (D-B) point when $Q_{1}$ (a part of $Q$) equals to zero, where the interaction can be probed without degeneracy between the constant DE equation of state (EoS).
\end{abstract}

\section{Introduction}
Astronomical observations of supernovae \citep{SN_1, SN_2}, cosmic microwave background radiation (CMB) \citep{CMB_1, CMB_2, planck2015} and large-scale structure \citep{method_2_another, r10} indicate that our universe is currently undergoing an accelerated expansion. Among the various observations, the Hubble parameter $H$ directly shows the expansion by its definition: $H=\dot{a}/a$, where $a$ is the cosmic scale factor and $\dot{a}$ is its change rate with respect to cosmic time \citep{H}. According to general relativity, $H$ depends on the constituents of the universe. Planck \citep{planck2015} shows that the present universe consists of approximately $69.1\%$ of DE, $25.9\%$ of DM, $4.9\%$ of baryon matter and a small amount of radiation. DE and DM are hypothetical form of energy and matter that spread throughout our universe. However, they can not be directly measured since little or no detectable radiations are emitted by themselves. Despite of this, they have obvious effects on the universe. DE causes the cosmic accelerated expansion due to its negative pressure and DM affects the cosmic evolution via its gravity. In the $\Lambda CDM$ model, the DE candidate is a simple cosmological constant and can be understood as a vacuum energy with an EoS of $w=-1$. This leads to a constant energy density of DE. Most observations can be well explained by this model, which has consequently been accepted as the standard cosmological paradigm. However, some problems exist and alternative theories are possible \citep{lcdm_problems}. Also, some observational results remain challenging. For example, the observational Hubble parameter data (OHD) of $H\left ( z = 2.34 \right ) = 222\pm 7 \mathrm{\left [km\, s^{-1}\, Mpc^{-1}  \right ]}$ ($H_{2.34}$) that obtained by BOSS and lies below the prediction of the $\Lambda CDM$ model \citep{BOSS}. Many studies relate this $H_{2.34}$ data to a dynamical DE - a straightforward way to modify the $\Lambda CDM$ model. Ref.~\citep{hints} shows the DE is evolving from SN and $f_{gas}$ data, and is consistent with $H_{2.34}$ data. Ref.~\citep{tension} presents that a dynamical DE can alleviate the tension between $H_{2.34}$ data along with low redshift $H(z)$ and the standard $\Lambda CDM$ model. Ref.~\citep{models} demonstrates tests of a variety of models that allow for the evolution of DE based on BAO, CMB and SN data. Ref.~\citep{fit} fits a dynamical DE including $H_{2.34}$ data. Some other related work can be found in Refs.~\citep{dynamical_DE_boss_1, dynamical_DE_boss_2}. In this paper, DE is also considered to be dynamical, but with a different physical meaning - the interaction between DE and DM.

Before going forward we need to give a brief introduction of the DE-DM interaction. Starting from the standard $\Lambda CDM$ model, the energy densities of DE and DM evolve as $\rho _{\texttt{DE}} = \rho _{\texttt{DE},0}\times a^{-3(1+w)}$ and $\rho _{\texttt{DM}} = \rho _{\texttt{DM},0}\times a^{-3}$ respectively. Here $\rho _{0}$ is the energy density present-day. The current ratio of DE to DM in energy density is $\rho _{\texttt{DE}}/\rho _{\texttt{DM}}\approx  2.67$ \citep{planck2015}, which leads to the "Coincidence Puzzle". That is, why do the energy densities of DE and DM evolve at considerably different rates as the universe expands but happen to be of the same magnitude right now? To alleviate or resolve this puzzle, some researchers introduce an assumption regarding an interaction between DE and DM. This is quite reasonable, because the nature of DE and DM is unknown and interaction is permitted in the field theory \citep{field theory}. In this situation, the energy densities of DE and DM are linked to each other and thereby both become dynamical, providing themselves abilities to be comparable and thus lessening the coincidence. Some studies are as follows: Refs.~\citep{field theory, interaction_positive, interaction_explain_deviation, db_based_on_perturbation, interaction_c5, interaction_c6} show that energy transferring from DE to DM can alleviate the coincidence problem. In addition, Refs.~\citep{interaction_explain_deviation, interaction_reconstruction_boss, interaction_boss, interaction_CMB_boss} study the interaction including $H_{2.34}$ data. Ref.~\citep{db_based_on_perturbation} investigates possible ways to break the degeneracy between interaction and  DE EoS or DM abundance based on the perturbation evolution of DE and DM. Refs.~\citep{interaction_effects_on_cosmological_parameters_1, interaction_effects_on_cosmological_parameters_2} discuss the effects of interaction on cosmological parameters. Refs.~\citep{interaction_CMB_boss, CMB_interaction, interaction_on_CMB} study the effects of interaction on CMB. Refs.~\citep{dynamical_system_1, dynamical_system_2, dynamical_system_3} discuss the interaction models based on dynamical system. And some other related studies can be found in Refs.~\citep{interaction_1, interaction_several_models_1, interaction_several_models_2, probe_interaction, detect_interaction, interaction_quintessence_coincidence_1, interaction_quintessence_coincidence_2, interaction_scaling, exact_interaction_solution, large_scale_interaction, interaction_effects_on_large_scale_structure, interaction_observation_1, interaction_observation_2}.

However, instead of the coincidence puzzle or other subjects, we focus on the interaction term $Q$ itself and explore its model-independent properties. After analyzing the structure of $Q$, a model-independent D-B point is found when $Q_{1}=0$. At this point, the interaction does not depend on the constant DE EoS $w$. This property provides an opportunity for probing the interaction without degeneracy between $w$ at the D-B point in theory and makes tighter constraints of the interaction nearby the D-B point in practice. The Gaussian Process (GP) and Monte Carlo Markov Chain (MCMC) methods \citep{GP_o1, GP_o2} are used to reconstruct needed quantities that are not based on cosmological models. We reconstruct the distribution of the D-B point with two covariance functions for comparison. A normal distribution with a mean value of $\overline{z}_{\texttt{D-B}}\approx 1.4026$ or $\overline{z}_{\texttt{D-B}}\approx 1.3659$ is obtained with Gaussian or Matern $(v = 9/2)$ covariance function respectively. Though the location of the D-B point depends on the Hubble parameter and its derivatives, its property for breaking degeneracy is cosmological-model-independent.

This paper is organized as follows. In Section~\ref{sec:Gaussian Process}, the GP method is introduced in Subsection~\ref{sec:Methodology}, and then $H$ as well as its derivatives are reconstructed from OHD in Subsection~\ref{sec:GP reconstruction}. In Section~\ref{sec:Interaction and the D-B point}, the interaction term is introduced into cosmic evolution equations in Subsection~\ref{sec:Interaction introduce}, and the properties of the D-B point is discussed in Subsection~\ref{sec:The D-B point}. Finally, we draw conclusions in Section~\ref{sec:Conclusions}.

\section{Gaussian Process}
\label{sec:Gaussian Process}
In this section the methodology of GP is introduced in the first subsection, and $H(z)$ along with its derivatives are reconstructed via GP as examples in the second subsection.
\subsection{Methodology}
\label{sec:Methodology}
We don't know about the nature of DE and DM, so their physical models are poorly motivated from fundamental understandings. There is a good alternative way to use model-independent reconstruction method for finding a high confidence region that traps the true theory. GP meets our requirement since it is a fitting method with no need for parameterization function in advance \citep{GP_o1, GP_o2}. In other words, it is not based on cosmological models. However, two additional assumptions are required. The first assumption is that any two points on the function to be fitted are correlated by a covariance function. The second one is that all function points obey a joint Gaussian distribution. 

Regarding the first assumption, the Hubble parameter function $H(z)$ is considered as an example. The correlation between any two points $H(z_{i})$ and $H(z_{j})$ is calculated by the covariance function $k(z_{i}, z_{j})$, where the "Gaussian Squared Exponential" is employed for simplicity in mathematics \citep{GP_o1, GP_o2} (An alternative covariance function is considered in Subsection~\ref{sec:The D-B point}):
\begin{align}
&k\left ( z_{i} ,z_{j}\right )=\sigma_{f}^{2}\times Exp \left (-\frac{\left ( z_{i}-z_{j} \right )^{2}}{2l^{2}} \right ). \label{eq:Gaussian_cf}
\end{align}
As we can see, the correlation is increasing when these two points are approaching each other. When they coincide, the maximum correlation $\sigma _{f}^{2}$ is obtained. $\sigma_{f}$ and $l$ are two hyperparameters, which do not specify the form of the covariance function. They are determined by using the maximum likelihood method and depend on the observational data only \citep{GP_o1, GP_o2}. 

The second assumption is that all Hubble parameter values form a joint Gaussian distribution:
\begin{align}
\begin{bmatrix}
H_{1}\\ 
H_{2}\\ 
\vdots\\
H_{n}\\ 
H_{*}
\end{bmatrix}\sim \mathcal{N}\begin{pmatrix}
\begin{bmatrix}
\mu_{1}\\ 
\mu_{2}\\ 
\vdots\\
\mu_{n}\\ 
\mu_{*}
\end{bmatrix} &,& \begin{bmatrix}
 k_{11}& k_{12} &\ldots&k_{1n}&k_{1*} \\ 
 k_{21}&k_{22}  &\ldots&k_{2n}& k_{2*}\\ 
 \vdots &\vdots&\vdots&\vdots &\vdots\\
 k_{n1}&k_{n2}  &\ldots&k_{nn}& k_{n*}\\ 
 k_{*1}&  k_{*2}&\ldots&k_{*n}& k_{**} \nonumber
\end{bmatrix}
\end{pmatrix}=\mathcal{N}\left ( \vec{\mu},\vec{K} \right ).
\end{align}
$H_{1}$, $H_{2}$ $\cdots$ $H_{n}$ are $n$ observational values, and $H_{*}$ is the value to be fitted at the redshift $z_{*}$. They form a $\left (n+1\right )$-dimensional joint Gaussian distribution. $\vec{\mu}$ is a priori mean function of the covariance and is assumed to be $\vec{0}$ \citep{GP_o1}. 

So far, the reconstructed value $H_{*}$ can be obtained. It is a Gaussian random variable with a mean value $\overline{H}_{*}$ and a standard deviation $\sigma _{H_{*}}$ as follows:
\begin{align}
&\overline{H}_{*}=\vec{k}_{*}\vec{k}_{n}^{-1} \left [ H_{1}, H_{2}\cdots H_{n}\right ]^{T}, \nonumber\\ 
&\sigma_{H_{*}}^{2}=k_{**}-\vec{k}_{*}\vec{k}_{n}^{-1}\vec{k}_{*}^{T}, \nonumber
\end{align}
where 
\begin{align}
&\vec{k}_{n} = \left [ k\left (z_{i},z_{j}  \right ) \right ] \left ( i,j = 1,2\cdots n \right ), \nonumber\\
&\vec{k}_{*}=\left [ k_{1*},k_{2*}\cdots k_{n*}  \right ],\; k_{**}= \sigma_{f}^{2}.\nonumber
\end{align}

When the observational error is considered, $\vec{k}_{n}$ needs to be changed as:
\begin{align}
\vec{k}_{\, with\; error}=\vec{k}_{n}+\vec{\delta} _{ij}\vec{\sigma}^{2} _{error}. \nonumber
\end{align}

The derivatives of $H_{*}$ can also be reconstructed. Its mean value and corresponding standard deviation are:
\begin{align}
&\overline{H}_{*}^{(m)}=\vec{k}_{*}^{(m)}\vec{k}_{n}^{-1} \left [ H_{1}, H_{2}\cdots H_{n}\right ]^{T}, \nonumber\\ 
&\sigma_{H_{*}^{(m)}}^{2}=k_{**}^{(m,m)}-\vec{k}_{*}^{(m)}\vec{k}_{n}^{-1}\vec{k}_{*}^{(m)T}, \nonumber
\end{align}
where $m$ denotes $m^{th}$ derivative with respect to $z_{*}$.

After the function $H$ and its derivatives are reconstructed, a composed function $F\left (z, H, {H}', {H}''\cdots  \right )$ may be considered next. In this situation, $H$ and its derivatives are generally correlated at the same redshift $z_{*}$:
\begin{align}
& Cov\left (H_{z_{*}}^{(i)} , H_{z_{*}}^{(j)} \right ) = k_{**}^{(i,j)} - \vec{k}_{*}^{(i)}\vec{k}_{n}^{-1}\vec{k}_{*}^{(j)T}, \label{eq:cov}
\end{align}
where $i$,$j$ denote the $i^{th}$, $j^{th}$ derivative with respect to $z_{*}$. Therefore, random sampling is needed for reconstruction. We employ the Monte Carlo sampling method to obtain one set of values of $H$, ${H}'$, ${H}''$ $\cdots$ at $z_{*}$ from a multivariate normal distribution, whose covariance function is Eq.(\ref{eq:cov}). Then every set can lead to one function value $F(z_{*})$. After repeating this for many times, the distribution of $F(z_{*})$ can be obtained. Performing this process at different redshifts, the reconstructed function $F(z)$ can be achieved. 

In particular, the study about the reliability of GP can be found in Ref.~\citep{reconstruction_interaction}. For more GP related work, please refer to Refs.~\citep{optimizing_GP, m92_1, m92_2, m92_3, gapp_2017_92, gapp_2018_92, gapp_2018}. 

\subsection{GP reconstruction}
\label{sec:GP reconstruction}
\begin{figure}
    \includegraphics[width=\columnwidth]{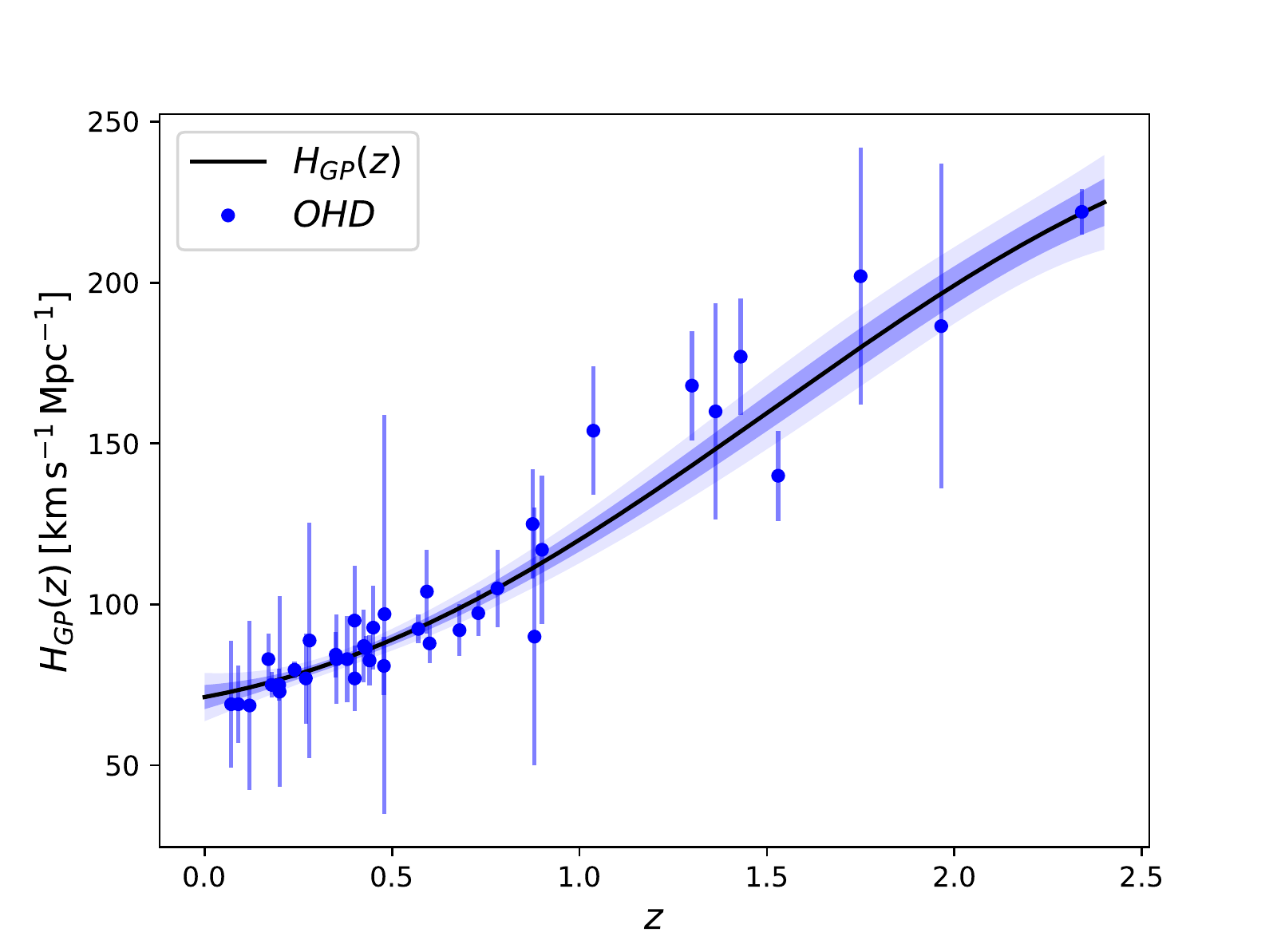}
    \caption{Reconstruction of $H_{\texttt{GP}}(z)$ via GP from OHD (Table~\ref{tab:Observational Hubble parameter data}). The blue points with errors are OHD. The black line is the GP reconstruction result $H_{\texttt{GP}}(z)$. The blue bands show the $1\sigma$  and $2\sigma$ regions of reconstruction.}

    \label{fig:Fitting of Observational Hubble parameter data}
\end{figure}

\begin{figure}
    \includegraphics[width=\columnwidth]{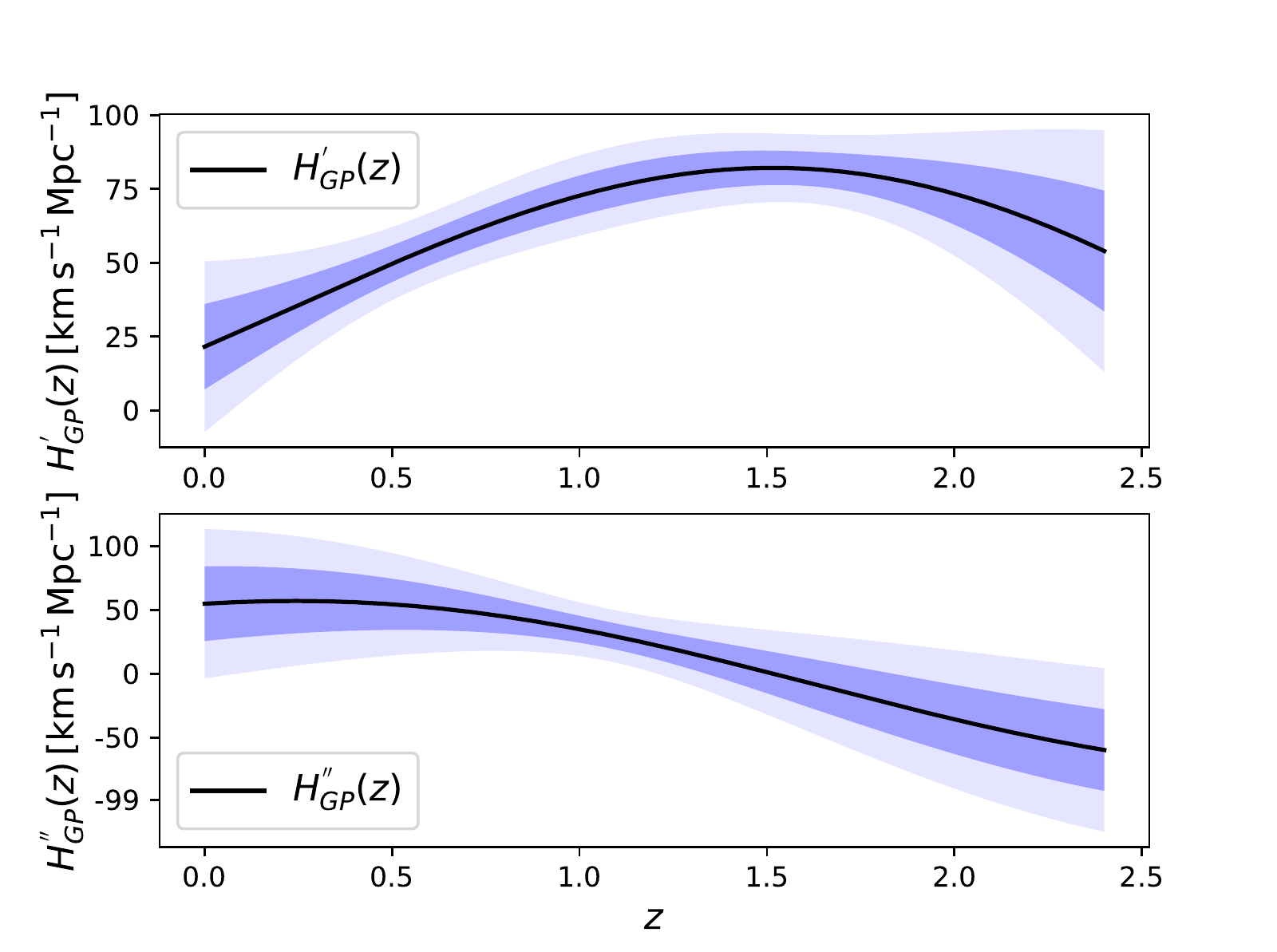}
    \caption{Upper panel: Reconstruction of ${H}'_{\texttt{GP}}(z)$ via GP from OHD (Table~\ref{tab:Observational Hubble parameter data}). Lower panel: Reconstruction of ${H}''_{\texttt{GP}}(z)$ via GP from OHD (Table~\ref{tab:Observational Hubble parameter data}). The blue bands show the $1\sigma$ and $2\sigma$ regions of reconstructions.}
    \label{fig:dH_d2H}
\end{figure}

As GP examples, $H(z)$ and its $1^{th}$ and $2^{th}$ derivatives are reconstructed from OHD (Table \ref{tab:Observational Hubble parameter data}). There are 38 data points in total and they are obtained from galaxy surveys via two methods \citep{two_methods_1, two_methods_2}: one is the differential galaxies age method, first proposed by Ref.~\cite{method_1}; the other is the radial BAO size method, discussed by Refs.~\cite{method_2_another, method_2} in the early time. The reconstruction of $H(z)$ is shown in Fig.~\ref{fig:Fitting of Observational Hubble parameter data}. The reconstructions of ${H}'(z)$ and ${H}''(z)$ are shown in Fig.~\ref{fig:dH_d2H}. A Python package "GaPP" including GP and MCMC methods is used. Details can be found in Refs.~\citep{GP_o1, GP_o2}. We run the program in Python 2.7.15 after modifying "== none" to "is none" in the source code.

\begin{table}
		\centering
		\begin{tabular}{|c|c|c|c|}\hline
			z&H$\; \mathrm{\left [km\: s^{-1}\: Mpc^{-1}  \right ]}$&Method&Reference\\\hline
0.0708&$69.0\pm 19.68$&1&\cite{H_1}\\
0.09&$69.0\pm 12.0$&1&\cite{H_2}\\
0.12&$68.6\pm 26.2$&1&\cite{H_1}\\
0.17&$83.0\pm 8.0$&1&\cite{H_3}\\
0.179&$75.0\pm 4.0$&1&\cite{H_4}\\
0.199&$75.0\pm 5.0$&1&\cite{H_4}\\
0.20&$72.9\pm 29.6$&1&\cite{H_1}\\
0.240&$79.69\pm 2.65$&2&\cite{H_5}\\
0.27&$77.0\pm 14.0$&1&\cite{H_3}\\
0.28&$88.8\pm 36.6$&1&\cite{H_1}\\
0.35&$84.4\pm 7.0$&2&\cite{H_6}\\
0.352&$83.0\pm 14.0$&1&\cite{H_4}\\
0.3802&$83.0\pm 13.5$&1&\cite{H_7}\\
0.4&$95.0\pm 17.0$&1&\cite{H_3}\\
0.4004&$77.0\pm 10.2$&1&\cite{H_7}\\
0.4247&$87.1\pm 11.2$&1&\cite{H_7}\\
0.43&$86.45\pm 3.68$&2&\cite{H_5}\\
0.44&$82.6\pm 7.8$&2&\cite{H_8}\\
0.4497&$92.8\pm 12.9$&1&\cite{H_7}\\
0.4783&$80.9\pm 9.0$&1&\cite{H_7}\\
0.48&$97.0\pm 62.0$&1&\cite{H_9}\\
0.57&$92.4\pm 4.5$&1&\cite{H_10}\\
0.593&$104.0\pm 13.0$&1&\cite{H_4}\\
0.6&$87.9\pm 6.1$&1&\cite{H_8}\\
0.68&$92.0\pm 8.0$&1&\cite{H_4}\\
0.73&$97.3\pm 7.0$&2&\cite{H_8}\\
0.781&$105.0\pm 12.0$&1&\cite{H_4}\\
0.875&$125.0\pm 17.0$&1&\cite{H_4}\\
0.88&$90.0\pm 40.0$&1&\cite{H_9}\\
0.9&$117.0\pm 23.0$&1&\cite{H_3}\\
1.037&$154.0\pm 20.0$&1&\cite{H_4}\\
1.3&$168.0\pm 17.0$&1&\cite{H_3}\\
1.363&$160.0\pm 33.6$&1&\cite{H_11}\\
1.43&$177.0\pm 18.0$&1&\cite{H_3}\\
1.53&$140.0\pm 14.0$&1&\cite{H_3}\\
1.75&$202.0\pm 40.0$&1&\cite{H_3}\\
1.965&$186.5\pm 50.4$&1&\cite{H_11}\\
2.34&$222.0\pm 7.0$&2&\cite{BOSS}\\\hline
		\end{tabular}
		\caption{OHD from galactic surveys. There are two methods used. In the above table, "1" represents the differential galaxies age method, and "2" represents the radial BAO size method.}
		\label{tab:Observational Hubble parameter data}
\end{table}

\section{Interaction and the D-B point}
\label{sec:Interaction and the D-B point}
In this section, the interaction term is introduced into the cosmic evolution equations in the first subsection, and properties of the D-B point are discussed in the second subsection.

\subsection{Interaction}
\label{sec:Interaction introduce}
It is necessary to know how the interaction term is introduced into cosmic evolution equations. Assuming a spatially flat Friedmann--Robertson--Walker universe, if there is no interaction between DE and DM, energy conservation is kept within each of them, namely:
\begin{align}
T_{\texttt{DE};\nu }^{\mu \nu} = 0,\;\;\;\;\;T_{\texttt{DM};\nu }^{\mu \nu} = 0, \label{eq:Tensor1}
\end{align}
where $T_{\texttt{DE}}^{\mu \nu }$  and $T_{\texttt{DM}}^{\mu \nu }$  are energy-momentum tensors of DE and DM separately. When interaction is introduced between them, Eq. (\ref{eq:Tensor1}) becomes:
\begin{align}
T_{\texttt{DE};\nu }^{\mu \nu} = F^{\mu},\;\;\;\;\;T_{\texttt{DM};\nu }^{\mu \nu} = -F^{\mu}, \label{eq:Tensor2}
\end{align}
here $F^{\mu }$ is the force from DM acting on DE. In this situation, DE and DM keep energy conservation as a whole \citep{Tensor}:
\begin{align}
\left (T_{\texttt{DE}}^{\mu \nu } + T_{\texttt{DM}}^{\mu \nu }  \right )_{;\nu } = 0. \nonumber
\end{align}
However, there exists energy transfer between DE and DM, making their energy densities both dynamical. The corresponding evolution equations with respect to redshift $z$ can be derived from Eq. (\ref{eq:Tensor2}), and are given as Eq. (\ref{eq:lez}) and Eq. (\ref{eq:lmz}). By combining these equations with the baryon density equation Eq. (\ref{eq:lbz}) and two Friedmann equations Eq. (\ref{eq:f1z}) and Eq. (\ref{eq:f2z}), we obtain a set of equations that describe the evolution of the universe after the cosmic recombination:
\begin{align}
&{\rho'_{\texttt{b}}}H\left ( 1+z \right )-3H\rho _{\texttt{b}}=0, \label{eq:lbz} \\
&{\rho'_{\texttt{DE}}}H\left ( 1+z \right )-3\left ( 1+w \right )H\rho _{\texttt{DE}}=Q, \label{eq:lez}\\
&{\rho'_{\texttt{DM}}}H\left ( 1+z \right )-3H\rho _{\texttt{DM}}=-Q, \label{eq:lmz} \\
&H^{2}=\left ( \rho _{\texttt{b}}+\rho _{\texttt{DM}}+\rho _{\texttt{DE}} \right )/3, \label{eq:f1z}\\
&{H}'H\left ( 1+z \right )=\left (\rho _{\texttt{b}}+\rho _{\texttt{DM}}+\left ( 1+w \right )\rho _{\texttt{DE}} \right)/2. \label{eq:f2z}
\end{align}
Here $'$ represents taking derivative with respect to the redshift $z$, and $a=1/(1+z)$ is used for variable transformation from $a$ to $z$. $\rho _{\texttt{b}}$, $\rho _{\texttt{DE}}$ and $\rho _{\texttt{DM}}$ are baryon, DE and DM energy densities respectively. The small amount of radiation is neglected. Dimensionless parameter $w$ is the constant DE EoS and $Q$ is the interaction term, which represents the energy transfer rate between DE and DM. If $Q<0$, the energy flows from DE to DM and vice versa. When $Q=0$ with $w=-1$, the standard $\Lambda CDM$ model is recovered. We set $8\pi G = 1$ for simplicity in this paper.

\subsection{The D-B point}
\label{sec:The D-B point}
The main results of our study are presented in this subsection, and the general content is outlined here. At the start, we obtain the interaction term $Q$ from cosmic evolution equations. After analyzing its structure, the D-B point is found when $Q_{1}=0$ and the value of $Q$ at the D-B point is most likely positive. After that, the feasibility of the D-B point in observation is discussed based on the error propagation between $Q$ and $w$. At last, distribution of the location of the D-B point is reconstructed and two different covariance functions are considered for comparison.

\begin{figure}
    \includegraphics[width=\columnwidth]{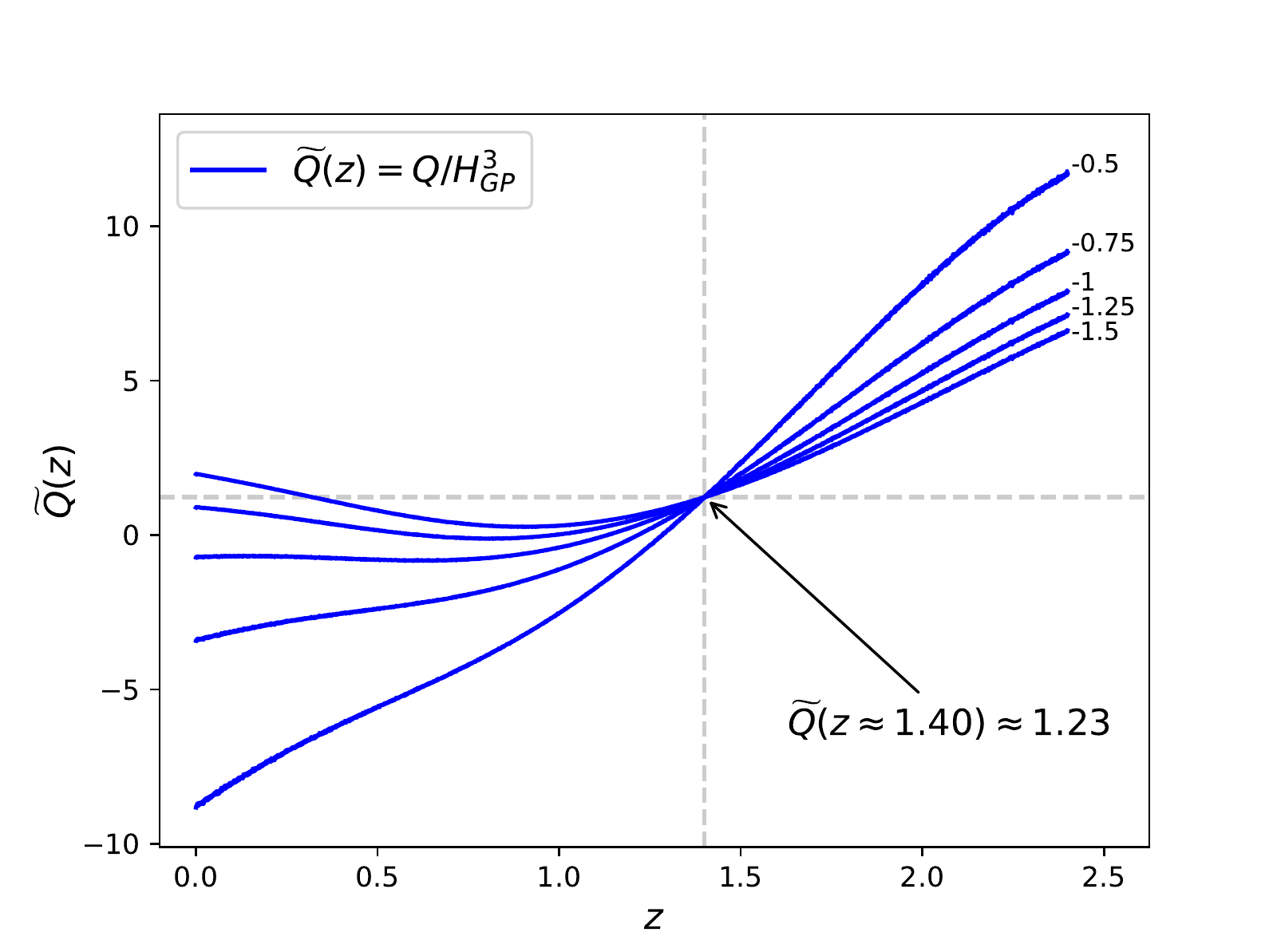}
    \caption{Curves of $\widetilde{Q}(z)=Q/H_{\texttt{GP}}^{3}$ with five different $w$ values, illustrating the D-B point at the crossover. Each blue curve is $\widetilde{Q}(z)$ with $w$ equals to the value noted at the top right. $\widetilde{Q}(z)$ is plotted here instead of $Q(z)$ to avoid curves being hard to distinguish at the lower value region. Each $Q(z)$ is reconstructed via GP and MCMC from OHD (Table~\ref{tab:Observational Hubble parameter data}). And $H_{\texttt{GP}}(z)$ is the mean value of the reconstructed $H(z)$ in Subsection~\ref{sec:GP reconstruction}. The gray dashed horizontal and vertical lines meet at the D-B point, which corresponds to $z\approx1.40$ and $Q( z_{\texttt{D-B}}) \approx 1.23H_{\texttt{GP}}^{3}( z_{\texttt{D-B}})$.}
    \label{fig:Q}
\end{figure}

\begin{figure}
    \includegraphics[width=\columnwidth]{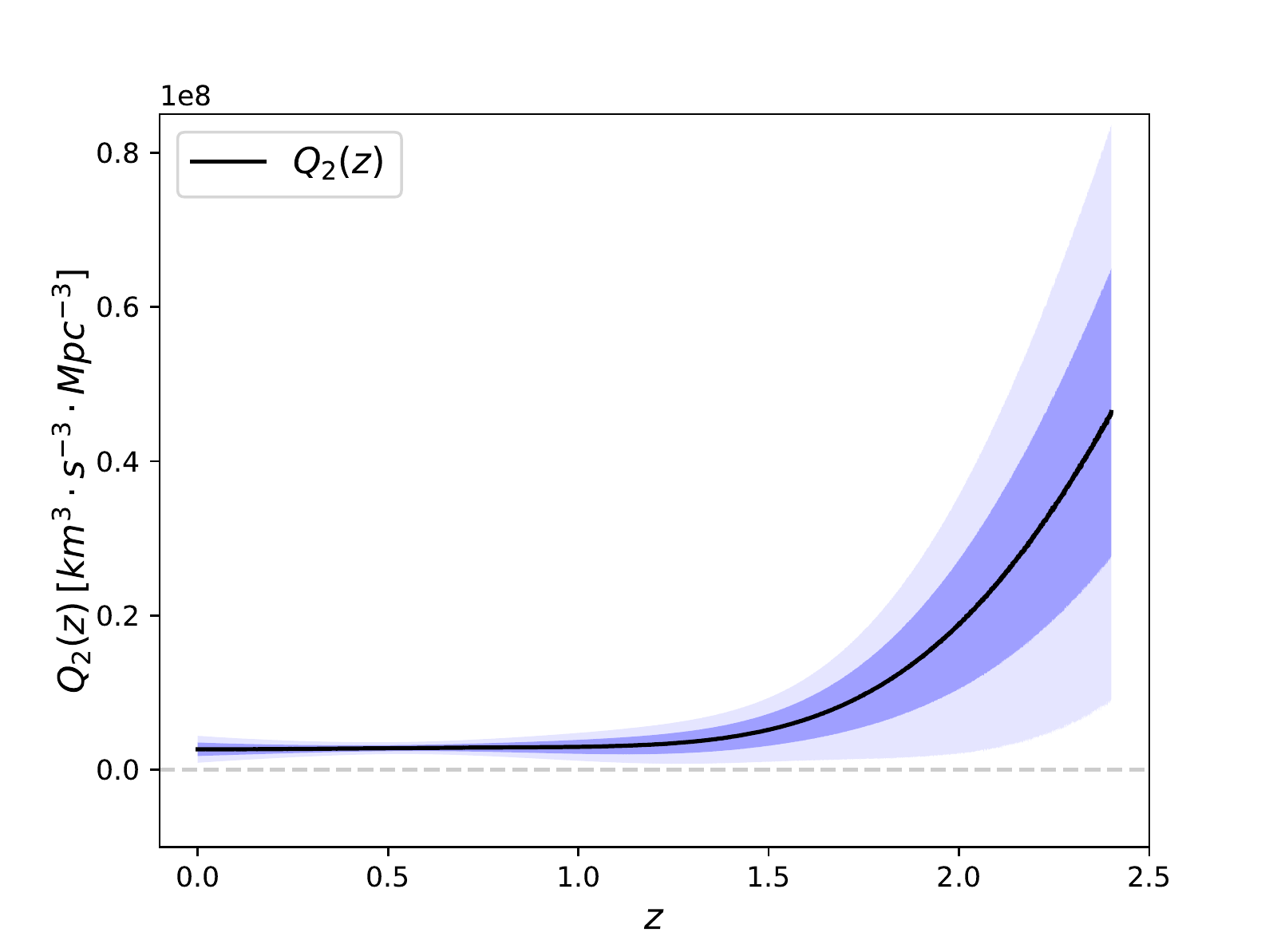}
    \caption{Reconstruction of $Q_{2}(z)$ via GP and MCMC from OHD (Table~\ref{tab:Observational Hubble parameter data}). The black line is $Q_{2}(z)$. The blue bands show $1\sigma$  and $2\sigma$ regions of reconstruction. The gray dashed horizontal line shows the value of zero. Note that the value of $Q_{2}(z)$ is greater than zero by more than $2\sigma$ at any redshift within $z\in[0, 2.4]$.}
    \label{fig:Q2}
\end{figure}

First of all, the theoretical expression of $Q$ is needed. We start from the DE energy density, which can be derived from Eq. (\ref{eq:f1z}) and Eq. (\ref{eq:f2z}) :
\begin{align}
\rho _{\texttt{DE}}=\frac{2{H}'H\left ( 1+z \right )-3H^{2}}{w}. \label{eq:rc1}
\end{align}
Then, by substituting Eq. (\ref{eq:rc1}) into Eq. (\ref{eq:lez}) we obtain the term $Q$, which is a sum of three terms, $Q_{1}$, $Q_{2}$ and $Q_{3}$:
\begin{eqnarray}
Q&=&\left [H\times \!\frac{2\left ( {H}'^{2}+H{H}'' \right )\left ( 1+z \right )^{2}\!-\!10H{H}'\left ( 1+z \right )\!+\!9H^{2}}{w}\right]\nonumber\\
&+&\left [H^{2}\times \frac{3w\left (3H-2{H}'\left ( 1+z \right )\right )}{w}\right]\nonumber\\
&+&\left [H\times \frac{{w}'\left ( 1+z \right )\left ( 3H^{2}-2H{H}'\left ( 1+z \right ) \right )}{w^{2}}\right]\nonumber\\
&=&Q_{1}+Q_{2}+Q_{3}. \label{eq:Q}
\end{eqnarray}
From a functional point of view, $Q$ is a function of $H(z)$, ${H}'(z)$, ${H}''(z)$, $z$ and $w$. So there is degeneracy between $Q$ and $w$ when $H(z)$ is determined. But there may exist some special redshifts where $Q$ is independent of $w$. To elaborate this, the structure of $Q$ is studied and the three terms are discussed in the order of $Q_{3}$, $Q_{2}$, and $Q_{1}$. First of all, $Q_{3}$ is equal to 0 since $w$ is assumed to be a nonzero constant in this paper. $Q_{2}$ is independent of $w$ since $w$ in the numerator and denominator cancel out. Finally, we focus on $Q_{1}$, whose numerator looks like a quadratic function with respect to $(1+z)$, but actually it is not. Considering $H(z)$ and its derivatives are indeed functions of $z$, $Q_{1}$ is a complex function of $z$ and $w$. But there exists special D-B points, which are determined by the redshift where $Q_{1}$ is equal to zero. At these points $Q$ is independent of $w$, in other words, the degeneracy between $Q$ and $w$ is broken. To illustrate the D-B point, five curves of $\widetilde{Q}(z, w)=Q/H_{\texttt{GP}}^{3}$ with different $w$ values are shown in Fig.~\ref{fig:Q}. The common crossover is the D-B point, which locates at $z\approx 1.40$. Each $Q(z, w)$ is reconstructed by using the GP and MCMC methods and the $H_{\texttt{GP}}(z)$ is the mean value of the reconstructed $H(z)$ in Subsection~\ref{sec:GP reconstruction}. $\widetilde{Q}$ is plotted instead of $Q$ to avoid curves being hard to distinguish at the lower value region. From Fig.~\ref{fig:Q}, we can know $Q( z_{\texttt{D-B}}) \approx 1.23H_{\texttt{GP}}^{3}( z_{\texttt{D-B}})$ - it is indeed positive. But considering its error of reconstruction, how far does it deviate from zero? It matters because of the following. When $Q( z_{\texttt{D-B}})\neq  0$, the interaction exists. However, when $Q( z_{\texttt{D-B}}) = 0$, the existence of the interaction is uncertain since dynamical $Q(z)$ may happen to be zero at the D-B point or there exists no interaction at all. Therefore, the D-B point loses most of its meaning when $Q( z_{\texttt{D-B}})$ is very close to zero. Fortunately, this situation corresponds to a very small probability. $Q_{2}$, which is equal to $Q$ at the D-B point and plotted in Fig.~\ref{fig:Q2}, shows a property of being greater than zero by more than $2\sigma$ at any redshift.  Here we use GP and MCMC methods for reconstruction. The results are the same when changing to different $w$ values since $Q_{2}$ is independent of it. So $Q( z_{\texttt{D-B}})$ is positive with a high probability. 

Although the existence of the D-B point is theoretically guaranteed, it is impractical to make measurements at this precise redshift because of  the unavoidable observational errors. However, the D-B point is still meaningful for breaking degeneracy between $Q$ and $w$ thereby making tighter constraints nearby the D-B point. This will be explained in the following.

\begin{figure}
    \includegraphics[width=\columnwidth]{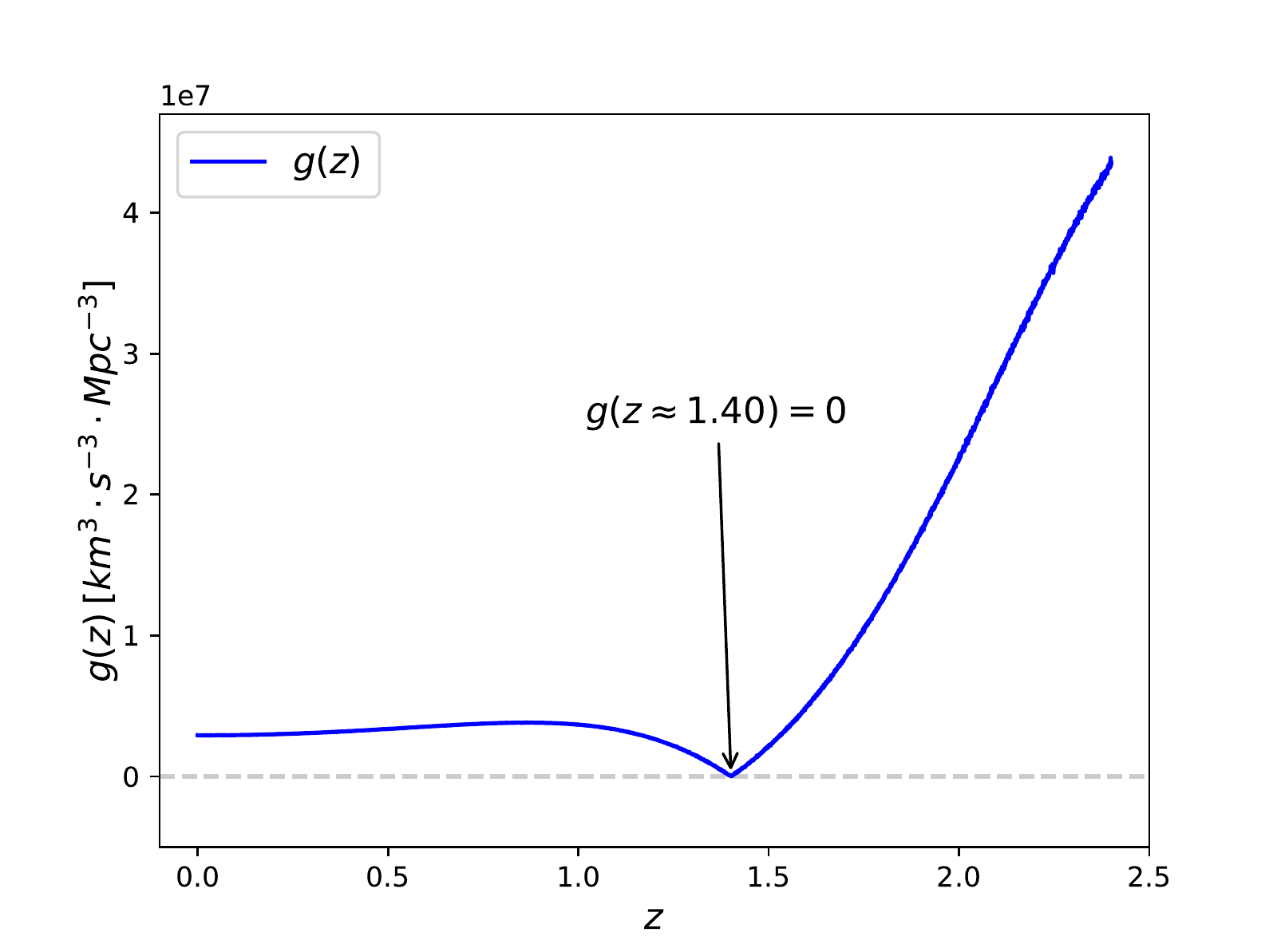}
    \caption{Amplification factor $g(z)$ of error propagation between $Q$ and $w$. The blue line is $g(z)$, which is reconstructed via GP and MCMC from OHD (Table~\ref{tab:Observational Hubble parameter data}). The gray dashed horizontal line shows the value of zero. Note that, the $g(z)$ reaches its minimum of zero at the D-B point and remains relatively small nearby. It means tighter constraints of the interaction can be made nearby the D-B point.}
    \label{fig:AF}
\end{figure}

In order to illustrate the feasibility of the D-B point in observation, we take derivative of $Q$ with respect to $w$. From Eq. (\ref{eq:Q}), we obtain the error propagation function as follows:
\begin{align}
\left |\Delta Q(z)  \right | = \left |\frac{{Q_{1}}(z)}{w}  \right |\times \left |\Delta w  \right | = g(z) \times \left |\Delta w  \right |,  \label{eq:err_p}
\end{align}
here $g(z)$ is an amplification factor that links the error of $Q(z)$ with that of $w$. The mean value of $g(z)$ is obtained by performing the GP and MCMC methods as shown in Fig.~\ref{fig:AF}. It is clear that near the D-B point, $g(z)$ is smaller than anywhere else. Though it is impossible to do cosmic measurements at this precise redshift, it is still helpful to make tighter constraints of the interaction nearby the D-B point in practice.

\begin{figure}
    \includegraphics[width=\columnwidth]{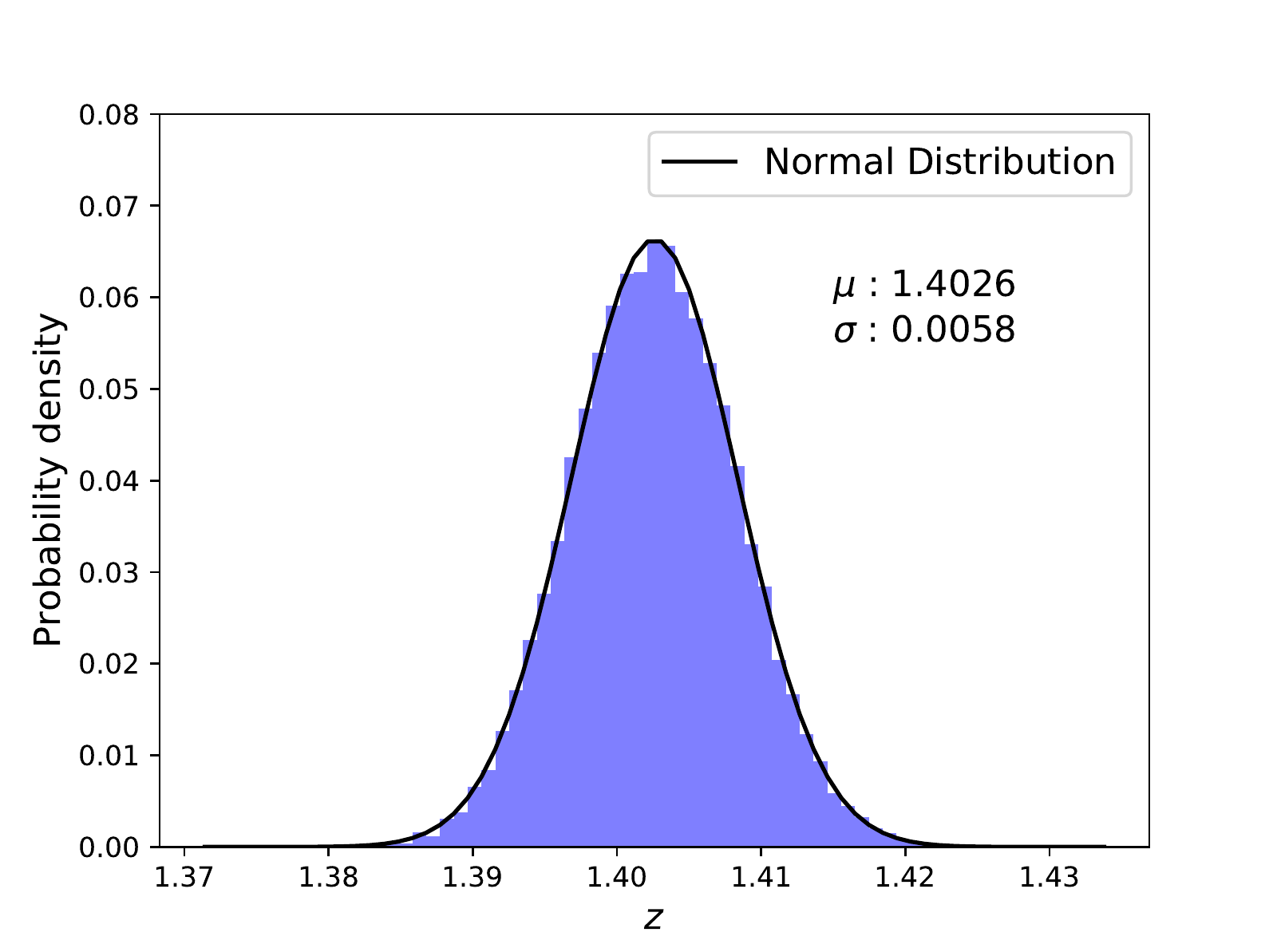}
    \caption{Distribution of the redshift of the D-B point. The black line is the probability density of a best-fit normal distribution. The mean value of the distribution is $\overline{z}_{\texttt{D-B}}\approx 1.4026$ and the standard deviation is $\sigma \approx 0.0058$.}
    \label{fig:z_dis}
\end{figure}

Now let's calculate the location of the D-B point. It is determined by the redshift where $Q_{1}$ is equal to zero. By using GP and MCMC methods, $Q_{1}$ can be reconstructed. And therefore the location of D-B point in the range of  $z\in [ 0,2.4]$ can be obtained: It belongs to a normal distribution with a mean value $\overline{z}_{\texttt{D-B}}\approx 1.4026$ and a standard deviation $\sigma\approx 0.0058$, which is shown in Fig.~\ref{fig:z_dis}. Here we apply the Gaussian covariance function as Eq. (\ref{eq:Gaussian_cf}). The optimized values of hyperparameters are ${\sigma_{f}\approx 157.62,l\approx 2.16}$. There are many choices of covariance functions. Ref.~\citep{optimizing_GP} Shows that the Matern $(v = 9/2)$ covariance function is a better form to obtain reliable results for supernovae data, given by:
\begin{eqnarray}
k(z_{i},z_{j}) &=&\sigma_{f}^{2}\times Exp(-\frac{3 |z_{i}-z_{j}|}{l})\times(1+\frac{3 |z_{i}-z_{j}|}{l}\nonumber\\
&+&\frac{27(z_{i}-z_{j})^{2}}{7l^{2}}+\frac{18(z_{i}-z_{j})^{3}}{7l^{3}}+\frac{27(z_{i}-z_{j})^{4}}{35l^{4}}).\nonumber
\end{eqnarray}
This form has been used in some other work \citep{m92_1, m92_2, m92_3, gapp_2017_92, gapp_2018_92}, leading the redshift of D-B point to a different normal distribution with $\overline{z}_{\texttt{D-B}}\approx 1.3659$ and $\sigma\approx 0.0064$. The optimized hyperparameters are ${\sigma_{f}\approx 164.11,l\approx 2.85}$ now. As we can see, the different choices of covariance function will affect the reconstructed $H(z)$ and its derivatives, hence affecting the location of the D-B point. Under certain conditions, the D-B points may even disappear. If zero is not a plausible value of $Q_{1}$, the D-B point no longer exists.

\section{Conclusions}
\label{sec:Conclusions}
We studied the interaction term $Q$ in cosmological models with interaction between DE and DM. Its structure shows that when $Q_{1}=0$ and the DE EoS $w$ is assumed to be constant, there exists a model-independent D-B point, where the degeneracy between $Q$ and $w$ is broken, thus the interaction can theoretically be probed. In order to explore model-independent properties of $Q$, the GP and MCMC methods are used to reconstruct the needed quantities that are not based on cosmological models. To illustrate the D-B point, functions $Q(z)$ with different $w$ values are reconstructed. The common crossover depicted in Fig.~\ref{fig:Q} is the D-B point, where $Q(z_{\texttt{D-B}})$ does not depend on the constant DE EoS $w$. And the value of $Q(z_{\texttt{D-B}})$ is greater than zero by more than $2\sigma$. Therefore, the D-B point is meaningful in theory. Though it is impossible to do cosmic measurements at this precise redshift due to unavoidable observational errors, it is still helpful to make tighter constraints of the interaction nearby the D-B point in practice. Lastly, we reconstructed the distributions of the D-B point with two covariance functions for comparison. A normal distribution with a mean value of $\overline{z}_{\texttt{D-B}}\approx 1.4026$ and a standard deviation $\sigma\approx 0.0058$ or $\overline{z}_{\texttt{D-B}}\approx 1.3659$ and $\sigma\approx 0.0064$ is obtained with Gaussian or Matern $(v = 9/2)$ covariance function respectively. Though the location of the D-B point depends on the Hubble parameter and its derivatives, its property for breaking degeneracy is cosmological-model-independent.

We thank Y. L. Li for useful discussions. Z. Zhou thanks Y. Qian et al. for suggestions of English language revising. We also thank the reviewer for very detailed comments, which offer great help for improving this paper.


\begin{thebibliography}{75}

\bibitem{SN_1}
A. G. Riess, A. V. Filippenko and P. Challis et al., Astron. J. \textbf{116}, 1009 (1998)
\href{https://arxiv.org/abs/astro-ph/9805201}{arXiv:astro-ph/9805201}
\bibitem{SN_2}
S. Perlmutter, G. Aldering and G. Goldhaber et al., Astrophys. J. \textbf{517}, 565 (1999)
\href{https://arxiv.org/abs/astro-ph/9812133}{arXiv:astro-ph/9812133}
\bibitem{CMB_1}
D. N. Spergel, R. Bean and O. Dore et al., Astrophys. J. Suppl. \textbf{170}, 377 (2007)
\href{https://arxiv.org/abs/astro-ph/0603449}{arXiv:astro-ph/0603449}
\bibitem{CMB_2}
G. Hinshaw, D. Larson and E. Komatsu et al., Astrophys. J. Suppl. \textbf{208}, 19 (2013)
\href{https://arxiv.org/abs/1212.5226}{arXiv:1212.5226 [astro-ph.CO]}
\bibitem{planck2015}
Planck Collaboration et al., Astron. Astrophys. \textbf{594}, A13 (2016)
\href{https://arxiv.org/abs/1502.01589v3}{arXiv:1502.01589v3 [astro-ph.CO]}
\bibitem{method_2_another}
D. J. Eisenstein, I. Zehavi and D. W. Hogg et al., Astrophys. J. \textbf{633}, 560 (2005)
\href{https://arxiv.org/abs/astro-ph/0501171}{arXiv:astro-ph/0501171}
\bibitem{r10}
W. J. Percival, B. A. Reid and D. J. Eisenstein et al., Mon. Not. Roy. Astron. Soc. \textbf{401}, 2148 (2010)
\href{https://arxiv.org/abs/0907.1660}{arXiv:0907.1660 [astro-ph.CO]}
\bibitem{H}
C. Ma and T. J. Zhang, Astrophys. J. \textbf{730}, 74 (2011)
\href{https://arxiv.org/abs/1007.3787}{arXiv:1007.3787v2 [astro-ph.CO]}
\bibitem{lcdm_problems}
P. Bull, Y. Akrami and J. Adamek et al., Phys. Dark Univ. \textbf{12}, 56 (2016)
\href{https://arxiv.org/abs/1512.05356}{arXiv:1512.05356v2 [astro-ph.CO]}
\bibitem{BOSS}
T. Delubac, J. E. Bautista and N. G. Busca et al., Astron. Astrophys. \textbf{574}, A59 (2015)
\href{https://arxiv.org/abs/1404.1801}{arXiv:1404.1801 [astro-ph.CO]}
\bibitem{hints}
V. H. Cardenas, Phys. Lett. B \textbf{750}, 128 (2015)
\href{https://arxiv.org/abs/1405.5116v2}{arXiv:1405.5116v2 [astro-ph.CO]}
\bibitem{tension}
V. Sahni, A. Shafieloo and A. A. Starobinsky, Astrophys. J. \textbf{793}, L40 (2014)
\href{https://arxiv.org/abs/1406.2209v3}{arXiv:1406.2209v3 [astro-ph.CO]}
\bibitem{models}
E. Aubourg, S. Bailey and J. E. Bautista et al., Phys. Rev. D \textbf{92}, 123516 (2015)
\href{https://arxiv.org/abs/1411.1074}{arXiv:1411.1074v3 [astro-ph.CO]}
\bibitem{fit}
G. B. Zhao, M. Raveri and L. Pogosian et al., Nat. Astron. \textbf{1}, 627 (2017)
\href{https://arxiv.org/abs/1701.08165}{arXiv:1701.08165}
\bibitem{dynamical_DE_boss_1}
A. G. Valent, J. Sola and S. Basilakos, JCAP \textbf{01}, 004 (2015)
\href{https://arxiv.org/abs/1409.7048v3}{arXiv:1409.7048v3 [astro-ph.CO]}
\bibitem{dynamical_DE_boss_2}
R. Y. Guo and X. Zhang, Eur. Phys. J. C \textbf{76}, 163 (2016)
\href{https://arxiv.org/abs/1512.07703v4}{arXiv:1512.07703v4 [astro-ph.CO]}
\bibitem{field theory}
S. Micheletti, E. Abdalla and B. Wang, Phys. Rev. D \textbf{79}, 123506 (2009)
\href{https://arxiv.org/abs/0902.0318v4}{arXiv:0902.0318v4 [gr-qc]}
\bibitem{interaction_positive}
C. Feng, B. Wang and E. Abdalla et al., Phys. Lett. B \textbf{05}, 066 (2008)
\href{https://arxiv.org/abs/0804.0110v2}{arXiv:0804.0110v2 [astro-ph]}
\bibitem{db_based_on_perturbation}
J. H. He, B. Wang and E. Abdalla, Phys. Rev. D \textbf{83}, 063515 (2011)
\href{https://arxiv.org/abs/1012.3904v3}{arXiv:1012.3904v3 [astro-ph.CO]}
\bibitem{interaction_c5}
E. Abdalla, L. L. Graef and B. Wang, Phys. Lett. B \textbf{726}, 786 (2013)
\href{https://arxiv.org/abs/1202.0499}{arXiv:1202.0499v2 [gr-qc]}
\bibitem{interaction_c6}
A. A. Costa, X. D. Xu and B. Wang et al., JCAP \textbf{01}, 028 (2017)
\href{https://arxiv.org/abs/1605.04138v2}{arXiv:1605.04138v2 [astro-ph.CO]}
\bibitem{interaction_explain_deviation}
E. G. M. Ferreira, J. Quintin and A. A. Costa et al., Phys. Rev. D \textbf{95}, 043520 (2017)
\href{https://arxiv.org/abs/1412.2777}{arXiv:1412.2777v4 [astro-ph.CO]}
\bibitem{interaction_reconstruction_boss}
Y. T. Wang, G. B. Zhao and D. Wands et al., Phys. Rev. D \textbf{92}, 103005 (2015)
\href{https://arxiv.org/abs/1505.01373v2}{arXiv:1505.01373v2 [astro-ph.CO]}
\bibitem{interaction_boss}
S. Pan and G. S. Sharov, Mon. Not. Roy. Astron. Soc. \textbf{472}, 4 (2017)
\href{https://arxiv.org/abs/1609.02287v2}{arXiv:1609.02287v2 [gr-qc]}
\bibitem{interaction_CMB_boss}
L. Santos, W. Zhao and E. G. M. Ferreira et al., Phys. Rev. D \textbf{96}, 103529 (2017)
\href{https://arxiv.org/abs/1707.06827v3}{arXiv:1707.06827v3 [astro-ph.CO]}
\bibitem{interaction_effects_on_cosmological_parameters_1}
L. Amendola, G. C. Campos and R. Rosenfeld, Phys. Rev. D \textbf{75}, 083506 (2007)
\href{https://arxiv.org/abs/astro-ph/0610806v2}{arXiv:astro-ph/0610806v2}
\bibitem{interaction_effects_on_cosmological_parameters_2}
J. H. He and B. Wang, JCAP \textbf{0806}, 010 (2008)
\href{https://arxiv.org/abs/0801.4233}{arXiv:0801.4233 [astro-ph]}
\bibitem{CMB_interaction}
D. Pavon, S. Sen and W. Zimdahl, JCAP \textbf{05}, 009 (2004)
\href{https://arxiv.org/abs/astro-ph/0402067v1}{arXiv:astro-ph/0402067v1}
\bibitem{interaction_on_CMB}
J. H. He, B. Wang and P. Zhang, Phys. Rev. D \textbf{80}, 063530 (2009)
\href{https://arxiv.org/abs/0906.0677v2}{arXiv:0906.0677v2 [gr-qc]}
\bibitem{dynamical_system_1}
B. Gumjudpai, T. Naskar and M. Sami et al., JCAP \textbf{0506}, 007 (2005)
\href{https://arxiv.org/abs/hep-th/0502191}{arXiv:hep-th/0502191v2}
\bibitem{dynamical_system_2}
X. Zhang, Mod. Phys. Lett. A \textbf{20}, 2575 (2005)
\href{https://arxiv.org/abs/astro-ph/0503072v2}{arXiv:astro-ph/0503072v2}
\bibitem{dynamical_system_3}
M. S. Berger and H. Shojaei, Phys. Rev. D \textbf{73}, 083528 (2006)
\href{https://arxiv.org/abs/gr-qc/0601086}{arXiv:gr-qc/0601086v3}
\bibitem{interaction_1}
G. R. Farrar and P. J. E. Peebles, Astrophys. J. \textbf{604}, 1 (2004)
\href{https://arxiv.org/abs/astro-ph/0307316v2}{arXiv:astro-ph/0307316v2}
\bibitem{interaction_several_models_1}
Z. K. Guo, N. Ohta and S. Tsujikawa, Phys. Rev. D \textbf{76}, 023508 (2007)
\href{https://arxiv.org/abs/astro-ph/0702015v3}{arXiv:astro-ph/0702015v3}
\bibitem{interaction_several_models_2}
L. L. Honorez, B. A. Reid and O. Mena et al., JCAP \textbf{09}, 029 (2010)
\href{https://arxiv.org/abs/1006.0877v2}{arXiv:1006.0877v2 [astro-ph.CO]}
\bibitem{probe_interaction}
K. Koyama, R. Maartens and Y. S. Song, JCAP \textbf{10}, 017 (2009)
\href{https://arxiv.org/abs/0907.2126v3}{arXiv:0907.2126v3 [astro-ph.CO]}
\bibitem{detect_interaction}
P. C. Ferreira, D. Pavon and J. C. Carvalho, Phys. Rev. D \textbf{88}, 083503 (2013)
\href{https://arxiv.org/abs/1310.2160v1}{arXiv:1310.2160v1 [gr-qc]}
\bibitem{interaction_quintessence_coincidence_1}
G. Mangano, G. Miele and V. Pettorino, Mod. Phys. Lett. A \textbf{18}, 831 (2003)
\href{https://arxiv.org/abs/astro-ph/0212518}{arXiv:astro-ph/0212518v1}
\bibitem{interaction_quintessence_coincidence_2}
L. P. Chimento, A. S. Jakubi and D. Pavaon et al., Phys. Rev. D. \textbf{67}, 083513 (2003)
\href{https://arxiv.org/abs/astro-ph/0303145}{arXiv:astro-ph/0303145v1}
\bibitem{interaction_scaling}
L. Amendola, M. Quartin and S. Tsujikawa et al., Phys. Rev. D \textbf{74}, 023525 (2006)
\href{https://arxiv.org/abs/astro-ph/0605488}{arXiv:astro-ph/0605488v1}
\bibitem{exact_interaction_solution}
G. Caldera-Cabral, R. Maartens and L. A. Urena-Lopez, Phys. Rev. D \textbf{79}, 063518 (2009)
\href{https://arxiv.org/abs/0812.1827v2}{arXiv:0812.1827v2 [gr-qc]}
\bibitem{large_scale_interaction}
Y. H. Li and X. Zhang, Phys. Rev. D \textbf{89}, 083009 (2014)
\href{https://arxiv.org/abs/1312.6328v3}{arXiv:1312.6328v3 [astro-ph.CO]}
\bibitem{interaction_effects_on_large_scale_structure}
R. J. F. Marcondes, R. C. G. Landim and A. A. Costa et al., JCAP \textbf{12}, 009 (2016)
\href{https://arxiv.org/abs/1605.05264}{arXiv:1605.05264v3 [astro-ph.CO]}
\bibitem{interaction_observation_1}
V. Salvatelli, N. Said and M. Bruni et al., Phys. Rev. Lett. \textbf{113}, 181301 (2014)
\href{https://arxiv.org/abs/1406.7297}{arXiv:1406.7297v2 [astro-ph.CO]}
\bibitem{interaction_observation_2}
R. C. Nunes, S. Pan and E. N. Saridakis, Phys. Rev. D \textbf{94}, 023508 (2016)
\href{https://arxiv.org/abs/1605.01712}{arXiv:1605.01712v2 [astro-ph.CO]}
\bibitem{Tensor}
V. Faraoni, J. B. Dent and E. N. Saridakis, Phys. Rev. D \textbf{90}, 063510 (2014)
\href{https://arxiv.org/abs/1405.7288v2}{arXiv:1405.7288v2 [gr-qc]}
\bibitem{GP_o1}
M. Seikel, C. Clarkson and M. Smith, JCAP \textbf{06}, 036 (2012)
\href{https://arxiv.org/abs/1204.2832}{arXiv:1204.2832v2 [astro-ph.CO]}
\bibitem{GP_o2}
M. Seikel, S. Yahya and R. Maartens et al., Phys. Rev. D \textbf{86}, 083001 (2012)
\href{https://arxiv.org/abs/1205.3431}{arXiv:1205.3431v2 [astro-ph.CO]}
\bibitem{reconstruction_interaction}
T. Yang, Z. K. Guo and R. G. Cai, Phys. Rev. D, \textbf{91}, 123533 (2015)
\href{https://arxiv.org/abs/1505.04443}{arXiv:1505.04443v2 [astro-ph.CO] }
\bibitem{optimizing_GP}
M. Seikel and C. Clarkson, arXiv:1311.6678v1 [astro-ph.CO]
\href{https://arxiv.org/abs/1311.6678v1}{arXiv:1311.6678v1 [astro-ph.CO]}
\bibitem{m92_1}
V. C. Busti, C. Clarkson and M. Seikel, Mon. Not. Roy. Astron. Soc. Lett. \textbf{441}, 1 (2014)
\href{https://arxiv.org/abs/1402.5429}{ arXiv:1402.5429v1 [astro-ph.CO]}
\bibitem{m92_2}
S. Yahya, M. Seikel and C. Clarkson et al., Phys. Rev. D \textbf{89}, 023503 (2014)
\href{https://arxiv.org/abs/1308.4099}{arXiv:1308.4099v2 [astro-ph.CO]}
\bibitem{m92_3}
V. C. Busti and C. Clarkson, JCAP \textbf{05}, 008 (2016)
\href{https://arxiv.org/abs/1505.01821v2}{arXiv:1505.01821v2 [astro-ph.CO]}
\bibitem{gapp_2017_92}
D. Wang and X. H. Meng, Phys. Rev. D \textbf{95}, 023508 (2017)
\href{https://arxiv.org/abs/1708.07750v1}{arXiv:1708.07750v1 [astro-ph.CO]}
\bibitem{gapp_2018_92}
M. J. Zhang and H. Li, Eur. Phys. J. C \textbf{78}, 460 (2018)
\href{https://arxiv.org/abs/1806.02981v1}{arXiv:1806.02981v1 [astro-ph.CO]}
\bibitem{gapp_2018}
E. Elizalde, M. Khurshudyan and S. Nojiri, arXiv:1809.01961v1 [gr-qc]
\href{https://arxiv.org/abs/1809.01961v1}{arXiv:1809.01961v1 [gr-qc]}
\bibitem{two_methods_1}
T. J. Zhang, C. Ma and T. Lan, AdAst, \textbf{2010}, 184284 (2010)
\href{https://arxiv.org/abs/1010.1307}{arXiv:1010.1307}
\bibitem{two_methods_2}
X. W. Duan, M. Zhou and T. J. Zhang, arXiv:160503947 [astro-ph]
\href{https://arxiv.org/abs/1605.03947}{arXiv:1605.03947 [astro-ph.CO]}
\bibitem{method_1}
R. Jimenez and A. Loeb, Astrophys. J. \textbf{573}, 37 (2002)
\href{https://arxiv.org/abs/astro-ph/0106145}{arXiv:astro-ph/0106145}
\bibitem{method_2}
C. Blake and K. Glazebrook, Astrophys. J. \textbf{594}, 665 (2003)
\href{https://arxiv.org/abs/astro-ph/0301632}{arXiv:astro-ph/0301632}
\bibitem{H_1}
C. Zhang, H. Zhang and S. Yuan et al., Res. Astron. Astrophys. \textbf{14}, 1221 (2014)
\href{https://arxiv.org/abs/1207.4541}{arXiv:1207.4541}
\bibitem{H_2}
R. Jimenez, L. Verde and T. Treu et al., Astrophys. J. \textbf{593}, 622 2003
\href{https://arxiv.org/abs/astro-ph/0302560}{arXiv:astro-ph/0302560}
\bibitem{H_3}
J. Simon, L. Verde and R. Jimenez, Phys. Rev. D \textbf{71}, 123001 (2005)
\href{https://arxiv.org/abs/astro-ph/0412269}{arXiv:astro-ph/0412269}
\bibitem{H_4}
M. Moresco, L. Verde and L. Pozzetti et al., JCAP \textbf{07}, 053 (2012)
\href{https://arxiv.org/abs/1201.6658}{arXiv:1201.6658}
\bibitem{H_5}
E. Gaztanaga, A. Cabre and L. Hui, Mon. Not. Roy. Astron. Soc. \textbf{399}, 1663 (2009)
\href{https://arxiv.org/abs/0807.3551}{arXiv:0807.3551 [astro-ph]}
\bibitem{H_6}
X. Xu, A. J. Cuesta and N. Padmanabhan et al., Mon. Not. Roy. Astron. Soc. \textbf{431}, 2834 (2013)
\href{https://arxiv.org/abs/1206.6732}{arXiv:1206.6732}
\bibitem{H_7}
M. Moresco, L. Pozzetti and A. Cimatti et al., JCAP \textbf{05}, 014 (2016)
\href{https://arxiv.org/abs/1601.01701}{arXiv:1601.01701 [astro-ph.CO]}
\bibitem{H_8}
C. Blake, S. Brough and M. Colless et al., Mon. Not. Roy. Astron. Soc. \textbf{425}, 405 (2012)
\href{https://arxiv.org/abs/1204.3674}{arXiv:1204.3674}
\bibitem{H_9}
D. Stern, R. Jimenez and L. Verde et al., JCAP \textbf{1002}, 008 (2010)
\href{https://arxiv.org/abs/0907.3149}{arXiv:0907.3149 [astro-ph.CO]}
\bibitem{H_10}
L. Samushia, B. A. Reid and M. White et al., Mon. Not. Roy. Astron. Soc. \textbf{439}, 3504 (2014)
\href{https://arxiv.org/abs/1312.4899}{arXiv:1312.4899 [astro-ph.CO]}
\bibitem{H_11}
M. Moresco, Mon. Not. Roy. Astron. Soc. Lett. \textbf{450}, L16 (2015)
\href{https://arxiv.org/abs/1503.01116}{arXiv:1503.01116 [astro-ph.CO]}

\end{thebibliography}
\end{document}